# Deep-blue light emitting diode based on defect variations of a 2D hybrid organic-inorganic low dimensional perovskite semiconductor

Ioanna Vareli[◊], Anastasia Vassilakopoulou[◊+], Ioannis Koutselas[◊]*

[◊]Materials Science Department, School of Natural Sciences, University of Patras, 26504 Greece

[+]Center for Technology Research & Innovation (CETRI) Ltd, Thessalonikis 1 Str.-Nicolaou Pentadromos Center, Limassol 3025 Cyprus

*ikouts@upatras.gr

**Abstract**

2D hybrid organic-inorganic semiconductors (HOIS) are low-cost, self-assembled semiconductors that have proven to be vital for novel optoelectronic devices due to their inherent stable excitons even at room temperature. 2D HOIS are chemically stable with respect to the 3D analogues, but their inherent anisotropy complicates their implementation in devices, including electroluminescence (EL) devices. We report on the synthesis, characterization, cathodoluminescence and EL of a new 2D HOIS, (4-fluoro phenethylamine-H)$_2$PbBr$_4$, and its defect variations (DVs). The latter, in contrast to the pristine material, allow for the first time the simple fabrication of a prototype blue EL device by coating DVs on a conducting anode substrate and using a Ga/In droplet cathode. DVs' optimization can lead to improved light emitting devices (LEDs) and the work here is expected to provide the incentive for utilizing HOIS DVs, as these are compatible with industrial scale synthesis. In contrast to the pristine iodine analogue, whose thin films readily provide EL, only the DVs or mixtures with $MoS_2$ nano-sized platelets provide EL, after careful deposition. The defects disrupt the planar non-conductive nature of the micrometric 2D HOIS platelets, usually aligned parallel to the substrate, thus, providing mechanisms for current flow, while the excitons' recombination half times are not affects by the defects. The usage of 4-fluoro-phenethylamine enhances the EL functionality of the DVs compared to using phenethylamine, due to the repulsion among the amine tails, leading to stable, weakly bound, decorated 2D inorganic sheets.

# 1. Introduction

Low Dimensional (LD) Hybrid Organic-Inorganic Semiconductors (HOIS) exhibit enticing optical and excitonic properties, due to the enhanced quantum and dielectric confinement of their inherent LD excitons. HOIS and especially their inorganic part, which is the semiconducting part, can attain the form of low dimensionality networks, from 3D to 0D, including intermediate dimensionalities such as the so called quasi-0D, quasi-1D and quasi-2D HOIS [1,2], while the superlattice formed out of the inorganic and organic parts bears the same active dimensionality. This natural, self-assembled and low-cost class of LD semiconductors exhibits useful optoelectronic properties comparable to those of the artificial semiconductor class and can be used, after some tuning, as photodetectors [3,4], radiation detectors [5], transistors [6], Light Emitting Diodes (LEDs) [7,8] or simply luminescent materials. LD HOIS exhibit excitonic states of increased binding energy ($E_b$) and oscillator strength ($f_{exc}$), while the associated excitonic Optical Absorption (OA) and accompanying Photoluminescence (PL) peak positions can be tuned by varying either the structure or their stoichiometry [9]. The increased excitonic binding energy ($E_b$) is mainly due to the dielectric mismatch of the alternating inorganic and organic subparts of the HOIS. The inorganic part composed of a metal halide unit network is usually semiconducting, while the organic part is formed from an appropriate self-assembled packing of electrooptically inactive protonated amines. In the 2D HOIS, which self-assemble as a 2D superlattice [10], the thickness of the repeating active layer is c.a. 0.6 nm, while that of the organic layer is about 1-3 nm, depending on the choice of the organic molecule [11,12,13], whose schematic representation has been laid down as inset in Figure 1. In the past, a wide set of combinations of inorganic and organic parameters have led to the synthesis of HOIS with a variety of properties and applications. Such examples are HOIS with controllable excitonic peaks covering the ultraviolet-visible region[1], thin film transistor gate materials comparable to amorphous Si [14] or HOIS exhibiting extremely interesting energy transfer optical phenomena [15]. Finally, the 3D and quasi-2D lead halide HOIS have been successfully employed for solar cell applications [16, 17, 18].

In this work, we report a new simple, low cost method for fabricating a 2D HOIS based excitonic LED, formed and operating at room temperature (RT). Initial results reported here show the advantages of the hybrid organic-inorganic lead halide perovskites in performing basic and applied research, specially in cases where stable 2D excitons are required at room temperature. Originally, 2D HOIS based LEDs were published in ref. [7] operating at 77K, while LEDs functioning at room temperature (RT) were reported in 2011 using oleylamine based HOIS [19], while in the same ref. the 4-fluorophenethylamine based HOIS were also mentioned as allowing the simple construction of LEDs operating at room temperature. In particular, the mentioned oleylamine-based HOIS, which were extremely well dried, had the advantage of being extremely resilient to humidity and water, however, the LEDs demanded in some cases high voltages, i.e. above 10V, in order to exhibit their Electroluminescence (EL) depending on the film thickness and the halogen content of the semiconductor. Also, the main EL peaks were broad and had significant contribution from strong Stokes shifted emission *wrt* to the excitonic OA peak, which, however, are also observed in 2D tin iodide based HOIS LEDs. The latter could be attributed to impurities in the organic amine leading to nanostructures with red shifted or defect-like excitonic states or even to organic-based broad emission mechanisms [20]. In HOIS, with multiple excitonic states, including defect-like states, energy is being transferred from higher energy excitonic states to those of lower energy. Lately, this has also been observed in energy transfer (or better funneling) LEDs based on such energy transfer capable HOIS, see for example *refs.* [21, 22] and references therein. Finally, specifically in the context of the particular 2D HOIS presented here, the defect variants (DVs) show EL capability when applied as thin films on conducting substrates simply by touching with a Ga/In droplet as counterelectrode, where the prinstine 2D HOIS do not function as such EL under the same preparation scheme.

## 2. Experimental

**2.1 Chemicals:** 4-Fluorophenethylamine (abbr. FpA, 98%), acetonitrile (abbr. AcN, CHROMASOLV® Plus, for HPLC, ≥99.9%), N,N'-Dimethylformamide (abbr. DMF, 99.8%), N-Methyl-2-pyrrolidone (abbr. nMP, anhydrous, 99.5% ), Hydrobromic acid (abbr. HBr, ACS reagent, ≥47.0%), Lead (II) bromide (99.999% trace metals basis), Gallium–Indium eutectic (≥99.99% trace metals basis), fluorine doped indium tin oxide (FTO) coated glass (square, surface resistivity 15-25 Ω/sq) were obtained from Sigma Aldrich and used as such with no further purification.

**2.2 Synthesis:** Synthetic route was followed as reported in ref. [23] and is analogous to syntheses reported in refs. [5, 24]. In a typical synthesis, 7 mmole of FpA was mixed, at 45°C, with 3.6 mmole of HBr (FpA-H denotes the protonated amine molecule); 4 mmole $PbBr_2$ was added in 3 ml of AcN and treated with 7.2 mmole HBr at 45°C. The second solution was slowly added, under mixing, to the first solution. Slow cooling allowed the formation of white crystals, which were dried extremely well for any remaining solvent. This sample will be denoted as **jvdv1**, whose crystals exhibit strong PL even under ambient room light, with a naked-eye visible characteristic blue or violet coloring. LEDs based on this defect variant, as well as on the pristine HOIS and its DVs, were fabricated by a simple doctor-blade method where 15μL of solution (80 mg of the 2D HOIS $(FpA-H)_2PbBr_4$ in 250μl DMF) was deposited on the FTO surface while it was heated at 50°C for a short time period.

Moreover, DVs have been synthesized using different precursor molar ratio (Λ) FpA: HBr: $PbBr_2$ than the one reported above; jvdv1 has Λ=7:10.8:4; sample **jv411** refers to synthesis from DMF with Λ=4:1:1, **jv421** refers to synthesis from DMF with Λ=4:2:2 while **jv221** was synthesized from DMF with Λ=2:2:1, which stands also for the pristine 2D compound. Sample **jvdv2** is as jvdv1 with the exception that $PbBr_2$ was added in 2 ml DMF. Assuming that the ideal material would be based on a precursor Λ ratio of 2:2:1, sample jvdv1 is expected to have some $PbBr_2$ impurities as it is amine deficient, while jv421 is expected to have packed extra amine molecules and finally jv411 is expected to have both amine and $PbBr_2$ impurities.

The FTO coated substrates were cleaned by immersion in piranha solution for 4 minutes and rinsed with 18MΩ water. No electron and hole injection layers were employed above or below the semiconductor layer, although this is possible and has been tested. LED's top contact was made using a Ga/In droplet electrode, however, other configurations can be used. The search for active LED materials was made using this simple configuration in order to avoid recombination effects arising from the electron/hole transport/blocking layers rather than from recombination within the HOIS layer. The HOIS-DMF solution used to prepare thin films for LEDs was also stored in closed transparent glass bottles under ambient air for one day to test for possible material oxidation, which was not observed in macroscopic optical test measurements but is speculated that it may occur. The degradation tests were made by forming spin coated films under the same conditions before and after the storage and comparing their absorbance spectra, as these were normalized to the maximum absorbance peak in the range of 300 – 800 nm.

$MoS_2$ nanoplatelets were synthesized by dissolving 16 mg $MoS_2$ in 1 ml N-Methyl-2-pyrrolidone (nMP) left in ultrasonic bath for 1 hour, the latter was centrifuged, and the supernatant solution was retained. This solution was added to 80 mg of jvdv1 dissolved in 250 μl DMF. The material resulted after drying this mixture is denoted as **jvdv1ms**. To facilitate simple, self-explanatory and yet short naming of the samples presented in this work, in Table 1 below summarizes the samples' short names, the related precursor molar ratio ($\Lambda$) as well as the solvent used for each sample.

**Table 1.** The precursor molar ratio ($\Lambda$) and the solvent used for each sample.

| Sample | $\Lambda$ := FpA: HBr: PbBr$_2$ | Solvent |
|--------|-------------------------------|---------|
| jvdv1  | 7: 10.8: 4                    | AcN     |
| jv411  | 4: 1: 1                       | DMF     |
| jv421  | 4: 2: 1                       | DMF     |
| jv221  | 2: 2: 1                       | DMF     |
| jvdv2  | 7: 10.8: 4                    | DMF     |
| jvdv1ms| 7: 10.8: 4                    | DMF/nMP/MoS$_2$ |

**2.3 Characterization:** X-ray diffraction (XRD) data were obtained on a Bruker D8 advance diffractometer equipped with a LynxEye detector and Ni filtered CuKa radiation. Optical Absorption (OA) spectra in the ultraviolet-visible spectral region were obtained on a Shimadzu 1650 spectrophotometer in the range of 200-800 nm. Samples were measured as thin films deposited on quartz substrates. Photoluminescence (PL) spectra were obtained on a Hitachi F-2500 FL spectrophotometer using the sample holder designed for solid samples, while in some cases thin films were mounted on substrates in a non-reflective geometry with respect to the excitation and detection slits, which were set at 2.5 nm and 2.5 nm, respectively. Video and images of the LEDs' operation were obtained by a standard computer camera, yet EL was easily visible by naked eye as well. Captured data were recorded for forward bias only, of about 6-12V, as reverse bias did not provide light emission. EL spectra were recorded by a modified θ-Metrisis optical profilometer using an optical fiber. CL images were recorded on a Leitz Wetzlar Orthoplan petrographic microscope with an attached Reliotron III Cathodoluminescence module and a custom-built module for light collection based on an Ocean Optics USB4000 device. SEM measurements were made with a Zeiss EVO-MA10 equipped with an Oxford EDS analyzer, where the electron beam accelerating voltages used were between 4.3-7 kV with currents of the order of up to 10nA for EDX measurements and 450 pA for imaging. The morphological study of the samples was observed without sputtering Au on the films, where charge accumulation was avoided by fast interlaced scanning and averaging techniques. STM and I-V measurements were made with a NanoScope III operating with a commercially acquired Pt/Ir tip under ambient atmosphere. Topographical measurements have been taken in both constant current and constant height mode for HOIS deposited on Au substrates and on FTO substrates, as well as to study the LEDs' rectifying behavior and perform scanning tunneling spectroscopic measurements (STS). For STS measurements the tip was fixed at the position determined by the set voltage ($V_{st}$) and set tunneling current ($I_{st}$) in order to obtain I-V characteristics curves, while the voltage was scanned from -5 to 5 V and in reverse to check for any hysteresis. The function dI/dV(V) was obtained by transforming the I(V) data, rather than using a lock-in amplifier to correlate the tunneling current to a fast oscillating bias. Bias voltages were applied to the sample with respect to the tip; positive current is defined as flowing from positive biased sample to tip. AFM measurements were performed with the same STM device, using standard AFM silicon nitride probe as supplied with the instrument.

Pico-second time-resolved fluorescence spectra were acquired using time correlated single photon-counting (TCSPC) with a Nano-Log spectrofluorometer (Horiba JobinYvon), using a laser diode as an excitation source (NanoLED, 375 nm) and a UV-Vis detector TBX-PMT series (250-850 nm) by Horiba JobinYvon. Lifetimes were evaluated with the Fluortools DecayFit Analysis Software as well as by custom written Mathematica Code to perform complex non-linear reconvolution fitting, with arbitrary types of functions. Instrumental Response Factor (IRF) spectra were measured using a Ludox aqueous solution, which have not been plotted with the decay spectra for simplicity reasons yet have automatically been taken into account for obtaining the models' parameters and half-times.

## 3. Results and Discussion

2D HOIS based on lead bromide and 4-fluorophenethylamine have been synthesized, characterized and implemented as a single layer LED device. Under appropriate reaction conditions, crystalline pristine 2D HOIS with chemical formula $(FpA-H)_2PbBr_4$ was produced, as well as defect variants of the previous were synthesized by using non-stoichiometric precursor molar ratios. 4-FpA has been used in the past by Mitzi for synthesizing low melting point temperature HOIS [14,18]. The final materials reported here comprise of white crystals, yielding strong photoluminescence observed as blue or blue-white by naked eye, when excited with a 404 nm laser, even though this laser excitation wavelength is below the band gap of the material yet higher than the exciton formation energy. Broad band white emission is sometimes due to some pure protonated amines.

Fig. 1 shows typical powder XRD patterns of the samples jvdv1 (a) and jv221 (b), plotted in log scale to show the details of the weak peaks. The XRD pattern in Fig.1a exhibits peaks at 5.33º, 10.64º, 15.98º, 21.34º, 26.76º, 32.24º, 37.80º, 43.44º while in Fig.1b the peaks are at 5.35º, 10.67º, 16.00º, 21.36º, 26.79º, 32.27º, 37.79º, 43.50º. These intense XRD peaks are practically identical, excluding some minor shifts due to sample thickness effects, yet their related small intensity peaks differ slightly. The low angle peak progression is due to the superlattice formed by the alternating organic and inorganic 2D layers, whose schematic representation is shown as inset in Fig.1. Solution of the unit cell crystal structure will be presented elsewhere for jv221; however, the cell clearly shows a c-axis repetition of 16.56Å, close the value reported for similar analogues HOIS [25] of 16.39Å and 16.65Å, due to the alternate stacking of the 2D organic layers and the inorganic 2D perovskite layers.

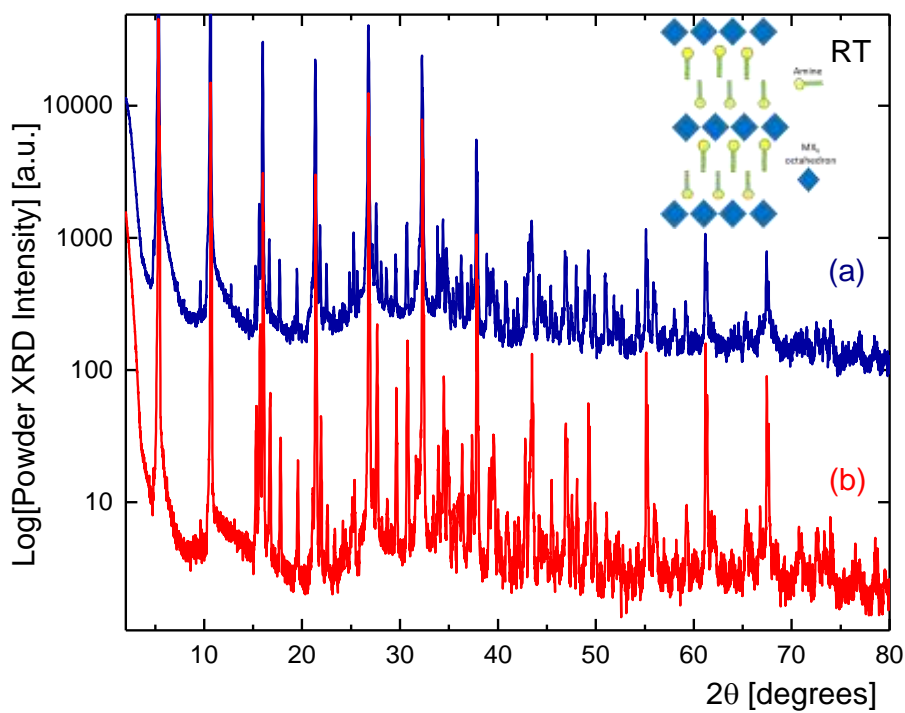

**Fig. 1** Powder XRD patterns of (a) jvdv1 and (b) jv221. Inset shows schematic representation of the 2D form of the pristine 2D HOIS.

The supplementary Figure S1, presents a comparison of the powder XRD patterns for the samples that were introduced in the experimental section as well as that of the pristine 2D HOIS and that of $PbBr_2$. All the 2D defect variants do show almost the same strong main XRD peaks as those of jvdv1, presented in Fig.1. It is interesting that although jvdv1 (shown in Fig.S1e), would have been expected to exhibit $PbBr_2$-like peaks, does show some spurious peaks resembling the pristine $PbBr_2$ pattern (Fig.S1f), yet the peaks have some considerable differentiation from those of $PbBr_2$. XRD patterns of the sample jvdv1ms can be found in the supplementary data as Fig.S2. It is interesting that the addition of nMP/$MoS_2$ does not induce any changes to the XRD pattern when compared to jvdv1, other than a small shift of the high angle orders, which implies that the long-range order of the perovskite is sustained in some cases, yet part of the perovskite has been altered.

Optical properties of the pristine 2D (FpA-H)$_2$PbBr$_4$ and its defect variants, were studied by means of OA and PL spectroscopy presented in Fig.2 left(L) and right(R), respectively, where the spectra indexes are the same for Fig.2L and Fig.2R. Fig.2L shows the OA spectra for various thin films deposited on quartz substrates and in some cases on the FTO substrates used for LEDs, while in Fig.2R the corresponding PL spectra have been acquired with $\lambda_{exc}$=300 nm, unless otherwise noted.

The lowest energy OA excitonic peaks occur for all samples at positions between 407 nm and 411 nm, with an onset of band gap absorption appearing at 375 nm that indicates an excitonic binding energy ($E_b$) of at least 260 meV; this increased value is due to the 2D nature of all the samples presented[1]. Only in the spectra of jv411 (Fig.2La) and jv421 (Fig.2Lb) the onset of band gap absorption appears to be at c.a. 388 nm, thus, $E_b$ is reduced to 150 meV. These observations point that jv411(Fig.2La), which is expected to have both $PbBr_2$ and amine impurities, appears to have decreased dielectric confinement, probably to impurity packing within the organic layer, thus, increasing its ε value. Also, it appears that sample jv421 (Fig.2Lb), which it would be expected to provide material similar to jv221 (Fig.2Lf), has slightly been altered compared to the latter probably due to the retaining of amine in its final structure as it was observed by the slightly shifted XRD peaks. Moreover, all OA spectra show a peak at 307 to 313 nm, which is due to absorption of the basic structural unit, the $PbBr_6^{2-}$ octahedron[1], which coincides with the OA peak of Pb-Br based 0D perovskites in the range of 307-325 nm[1]. Some variations of the peak position of the 0D OA peak are allowed due to the interaction of the octahedron with its environment.

Summarizing the OA results, it appears that all 2D HOIS variants exhibit an intense exciton state, with large binding energy, for these excitons are stable at room temperature. The OA spectra show that the 2D HOIS defect variant with the sharpest excitonic peak is jvdv1 as recrystallized from DMF (Fig.2Le), while jvdv1 as synthesized from AcN (Fig.2Lc) shows a broad OA peak which can be deconvoluted to two peaks at 400 nm and 408 nm.

PL excitonic peaks of the reported samples appear between 410 nm and 430 nm, while in some cases the PL peaks appear to be a convolution of two underlying peaks, which when combined yield relatively broad band emission.

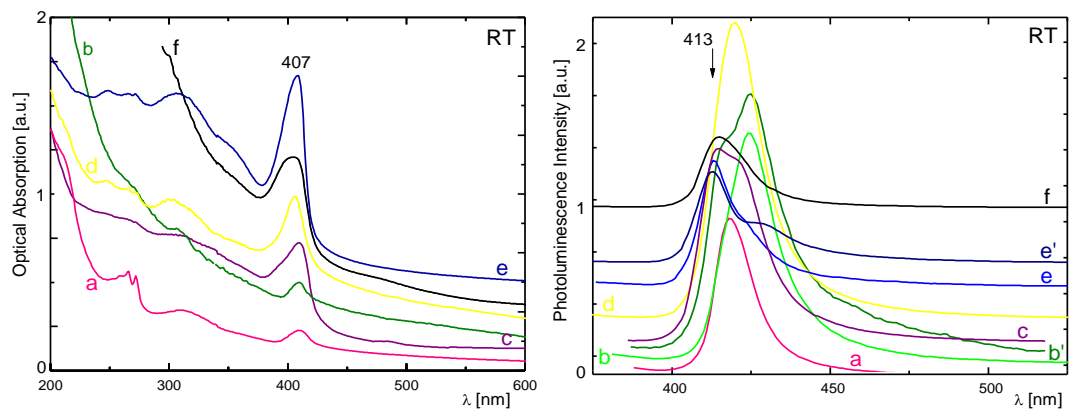

**Fig.2.** Left: OA spectra for (a) jv411, (b) jv421, (c) jvdv1, (d) jvdv2, (e) jvdv1 DMF/recrystallized and (f) jv221. Right: PL spectra ($\lambda_{exc}$= 300 nm, unless noted) for FTO substrates coated with (a) jv411, (b) jv421 ($\lambda_{exc}$=350 nm), (b′) jv421, (c) jvdv1, (d) jvdv2, (e) jvdv1 DMF/recrystallized, (e′) jvdv1 DMF/recrystallized ($\lambda_{exc}$=350 nm) and (f) jv221.

Sample jvdv1 as synthesized from AcN (Fig.2Rc) exhibits two PL peaks at 413 nm and 421 nm, yet when it is recrystallized from DMF the low energy emission shifts to 430 nm (Fig.2Re and Fig.2Re′), where the 430 nm emission is more pronounced when $\lambda_{exc}$= 350 nm. It is, thus, concluded that DMF induces a second phase in jvdv1 as evidenced in the PL spectrum, that is also present in the jvdv1 PL spectrum, in contrary to the OA spectra where DMF phase is not observed.

Also, jvdv2 (Fig.2Rd) which was synthesized in DMF has almost the same spectrum as jvdv1, nevertheless, jvdv2's PL emission peak is broader than of jvdv1 (Fig.2Rc) and centered at 420 nm. It should be noted that the PL peak of the $PbBr_2$ precursor, which is expected to coexist in samples jvdv1 and jvdv2, appears at 385 nm [1] and is not related to the peaks observed here centered at 421 nm or 430 nm.

Finally, spectra for the pristine 2D HOIS jv221 (f) shows that the excitonic peak of the pristine 2D compound appears at 407 nm for the OA spectrum and at 413 nm for the PL spectrum, the latter appears only slightly to be composed of a double peak. To summarize, the reported crystalline 2D HOIS exhibits an excitonic peak at c.a. 407 nm [1], as expected, while no other major phase/defect exist it due to its small PL Stokes shift. The other defect 2D variants are composed of an assortment of defects that do not appear, however, in an obvious fashion in the powder XRD patterns but do appear evidently in the OA and PL spectra. Finally, in the supplementary section it is possible to find the OA and PL spectra (Fig. S3) of jvdv1ms, which is jvdv1 mixed with $MoS_2$ nano-platelets. It is interesting that this sample shows an increased absorption at 320 nm, small absorption at the excitonic peak at 407 nm, but a pronounced PL signal at 417 nm. The shifted isolated octahedron peak appears at 320 nm, signifying the probable aggregation of $Pb_xBr_y$ cluster covered with stabilizing amine molecules and in any case, is due to isolated lead bromide clusters.

Although the pristine iodine analogue of the presented here 2D HOIS forms readily thin films that exhibit naked-eye visible EL[26], the bromine pristine 2D analogue does not provide EL films, not even at 77K, in the simple LED configuration described in the experimental section. Therefore, excluding phenomena related to band alignment and ease of carrier tunneling under external voltage, it would be assumed that the pristine 2D HOIS presented here ought to function as blue LED material. However, the pristine material, jv221, does not provide films with EL action, yet all the other reported variants do show EL, which is itself poses important questions. It has been theorized in ref. [26] that the fluorine atoms allow of repulsion among the organic counter-molecules defining the organic 2D planes due to the F-F interaction, thus, each inorganic layer is capped with organic material top and bottom with small interaction among the counter-organic molecules. This "free form" 2D set, organic/inorganic/organic, cannot disrupt the continuity of its inorganic component due to effects such as organic molecular vibrations, as these fluctuations cannot exist as a collective excitation. Moreover, the same F-F repulsion allows for the capability to grow more randomly oriented nanosized-structures that can adjust themselves so that current can flow from one 2D inorganic sheet to that of a neighboring particle's 2D inorganic sheet, circumventing the non-conducting path of the organic 2D planes.

More specifically, the films based on jvdv1 provided positive EL results, in terms of evaluating EL in-preparation simplicity, naked-eye observation, minimal set of control parameters (heating of the substrate, drying), film preparation, etc. In that aspect, the samples can be evaluated in the order as: jvdv1>jv411>jv421>jvdv1ms while jvdv2 and jv221 did not function at all as EL devices. It is important to note that other methods of LED fabrication may have provided EL even with the jv221 sample, yet in this specific LED fabrication that is used here, electrons are injected to the HOIS film via a Ga/In alloy electrode and holes are injected via the FTO electrode, while the device is prepared under standard lab conditions with no special precaution. In some cases, the contact of the Ga/In droplet at high voltage, c.a. 13V, gives rise to eye-visible white emission which may be related to the organic part emitting wide band radiation [27] but this in under investigation as this effect it has no preferential polarity and could be due to amine degradation.

The EL spectrum of a LED based on jvdv1 exhibits a peak at 417 nm as it is shown in Fig. 3, while the related figure inset presents the device's schematic and an image obtained during operation of the LED at 7.6 V. Usually, voltages between 5-10 V suffice for operation of the various blue LEDs. Video of LED's operation as well as snapshots from LED operation can be found as supplementary data (see Fig. S4).

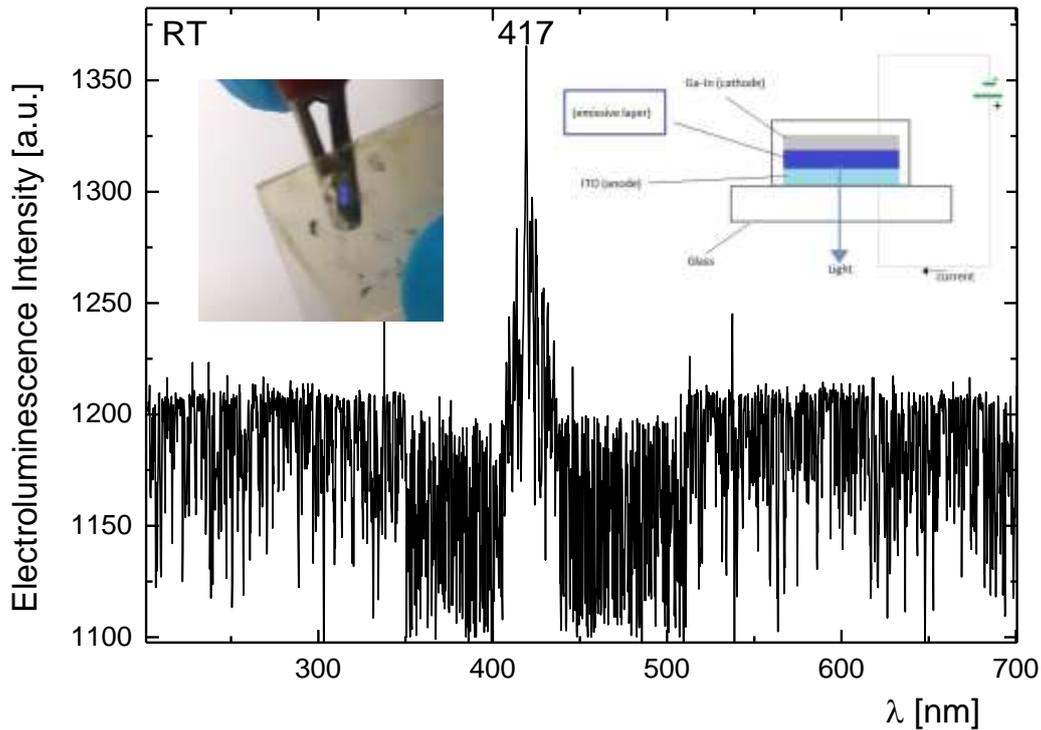

**Fig. 3.** EL spectra of LED fabricated by coating a heated FTO substrate with jvdv1 from heated DMF solution. Top left inset image shows the LED in operation; right inset shows device's schematic.

The EL peak is red shifted with respect to the corresponding OA peak of jvdv1 in Fig.2Lc by c.a. 10 nm as well as with respect to the high energy PL peak in Fig.2Rc. It is hypothesized that an interesting recombination process leads to the observed EL red shift, which is not related to phenomena attributed to the bulk of the material. Similar red shift has been observed and explained, for example in *refs*. [28,29], attributing it to various localized phenomena or high electric field polarization to which the interface band bending should be added.

To test whether bound excitons, defects or some sort of energy transfer [30, 31] is the origin of the red shift manifestations, a simple experiment was devised, as in [26]. It is possible for the EL radiation to appear peaked at lower energy than the exciton's OA peak, by a mechanism where EL radiation travelling through the sample undergoes absorption and re-emission, while finally the exiting emission appears to be redshifted. PL spectra of a jvdv1 film have been acquired in two different geometrical configurations at RT; the front face excitation and front face emission (FF) and the front face excitation with back face emission (BF), schematically represented as insets in Fig.4. It is observed that the film's PL is peaked at the same wavelengths for both BF and FF configurations for both the high energy and the low energy(defect) peaks, yet the intensity of the BF signal is notably decreased compared to that of the FF configuration.

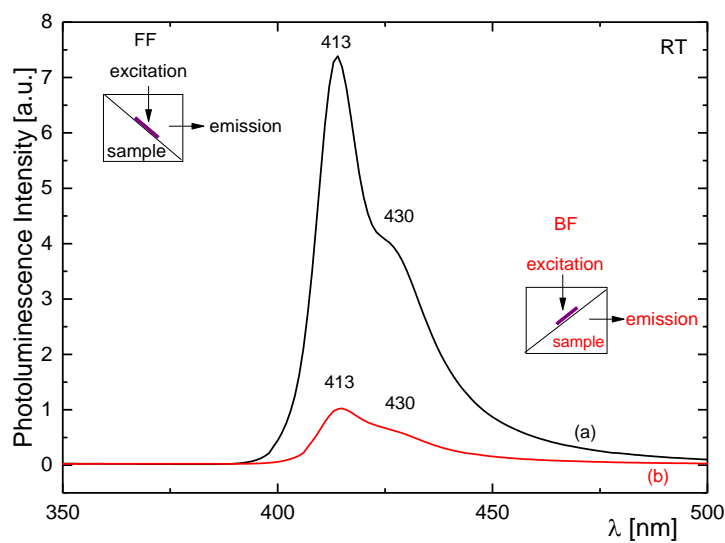

**Fig.4.** PL spectra ($\lambda_{exc}$=300 nm) of jvdv1 film deposited on a quartz plate, acquired in two geometries, front face excitation/emission (FF, a) front face excitation/back face emission (BF, b).

Therefore, it is suggested, since the BF configuration PL emission is practically identical to that of the FF configuration, that at room temperature there is no observed energy recycling mechanism responsible for the red shift of the EL since phenomena of absorption/re-emission or else the BF would have been red shifted [26]. Thus, the EL shifts are not due to any photon recycling mechanisms; it is necessary to note that the EL peak is expected to coincide with the OA peak, therefore, most probably defects are the origin of the red shifted peaks at lower wavelengths.

Herein, we have also exploited the potential of cathodoluminescence (CL) spectroscopy to verify the deep-blue emission of (4-fluorophenethylamine-H)$_2$PbBr$_4$ 2D HOIS defect variants. In Fig. 5 the CL image of jvdv1 (left) is presented along with the naked-eye visible luminescence upon excitation with 404 nm laser radiation (right). The intense scattered laser radiation at the center of Fig.5 (right) is registered as white by the camera and is not related to any white broad organic/defect PL emission.

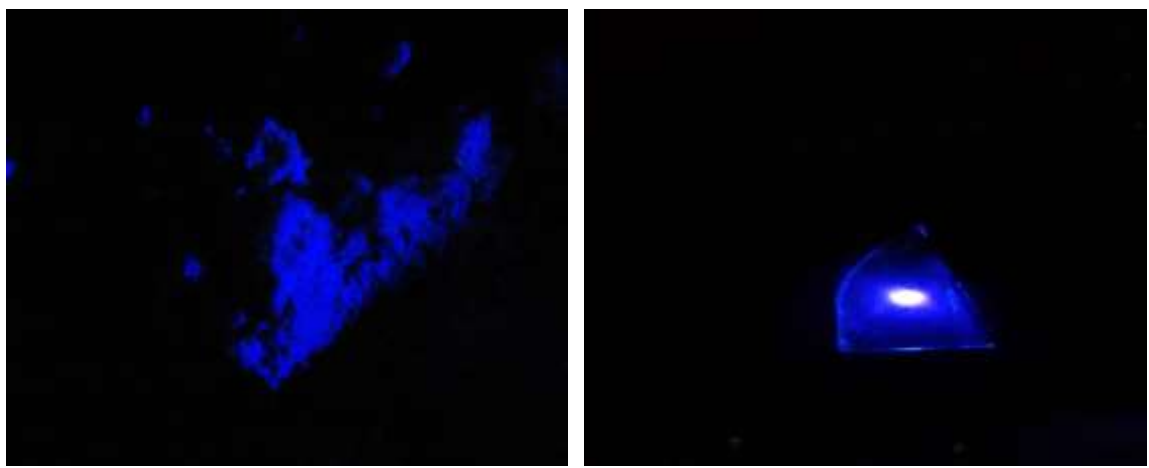

**Fig. 5.** CL image (left) of jvdv1 and under 404 nm excitation (right).

Fig. 6. presents SEM measurements carried out to investigate the surface morphology of our sample. A homogeneous distribution of platelets of $(FpA-H)_2PbBr_4$ crystals, with an EDX established molar ratio Pb:Br 1:4, is interrupted by needle-like phases, which appear to be identified by a molar ratio of at least 1:2.5 yet lower than 1:4. The needle formation of a 2D HOIS semiconductor has been observed and studied before in refs. [26,32], while the same phenomenon has been studied later in great detail in *ref.* [33] related to the properties of these "2D" needles.

Also, optical microscope images of jvdv2, jv411, jvdv1 and jv421 microcrystals can be found as supplementary data in the images Fig. S5, Fig. S6, Fig. S7, Fig. S8, respectively. It can be observed in these optical microscope images that the formation of rod-like microcrystals occurs in all samples but in the jvdv2, which is noted that it does not function as LED active material. Thus, it appears that simple LED device integration of these particular 2D HOIS defect variants is linked to the formation of rod-like phases.

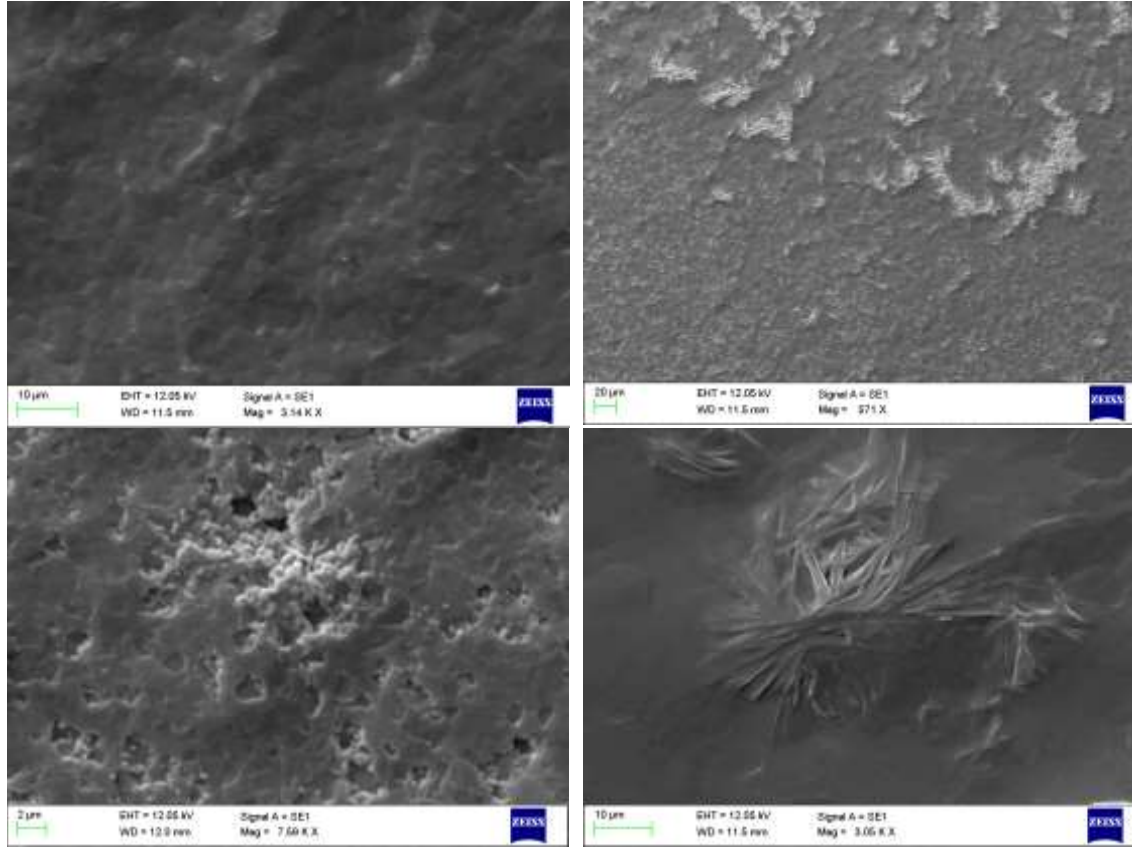

**Fig.6.** SEM images of jvdv1 as deposited on heated FTO substrates from heated DMF solution.

Surface topography was also evaluated using AFM and STM topography. Figure 7 shows 3D AFM and STM images for jvdv1 (Fig.7 top row) as deposited on FTO substrate; images acquired with $V_{st}$= -3.5 V and $I_{st}$=170 pA. In Fig.7 bottom row STM images of jvdv1 as deposited on Au substrate ($I_{st}$=107 pA, $V_{st}$=200 mV) are presented. The films which have been utilized as LEDs have a thickness of at least 150 nm, as suggested from the STM analysis and are not uniform in the nanoscale. On the other hand, the regions of the HOIS-coated Au substrates appear more uniform, which shows a need to address the HOIS film formation by other methods such as by applying heat and pressure as suggested in ref. [25].

The full acquired image area maximum surface roughness was calculated to be between 114 and 37 nm for the top and bottom row images, respectively, while the AFM large area analysis estimates a maximum surface roughness of 3 μm.

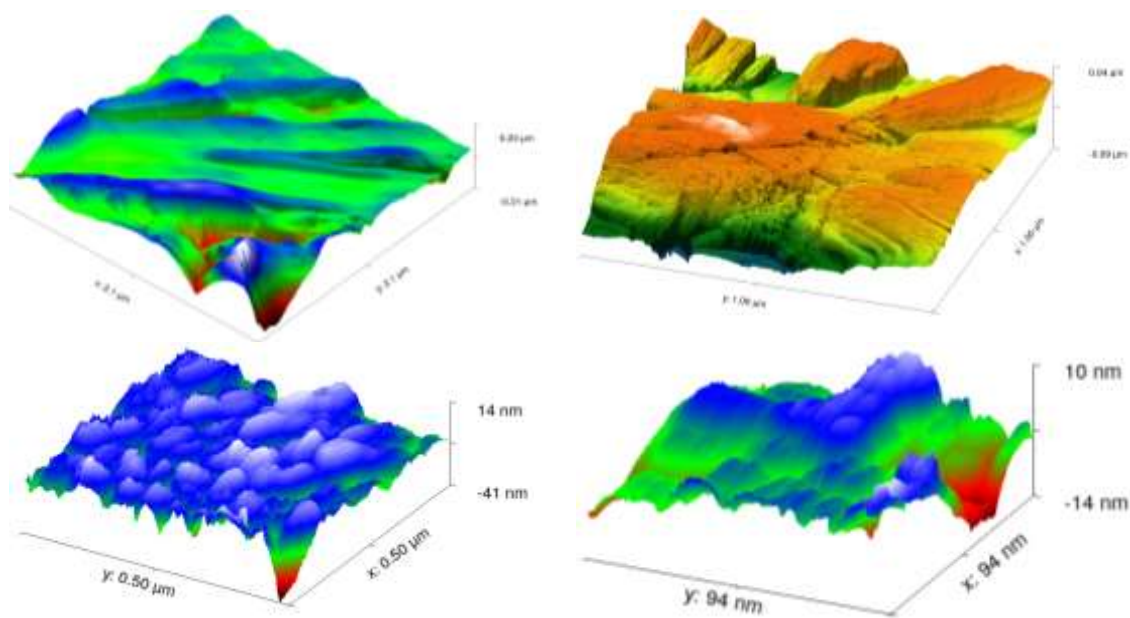

**Fig.7.** 3D AFM (top left) and STM (top right) images of jvdv1 deposited on heated FTO substrates from heated DMF solution. STM image was acquired at $I_{st}$= 174pA constant current mode and $V_{st}$= -3.5V. 3D STM images (bottom) of jvdv1 on Au substrate at $V_{st}$=0.2V and $I_{st}$=107pA at different resolutions.

I-V characteristic curves, as obtained by scanning the voltage difference between the substrate and the STM tip for the jvdv1/Au-substrate and the blue LED (jvdv1/FTO) can be found in Figs. 8 and 9, respectively. The graph regions where the I-V curves appear flat, at high absolute voltage values, are due to the saturation of the STM's A/D converter. Similar measurements were made for the bare Au and FTO surfaces. In particular, the bare Au surface provided the spectra (a) and (a′) for I(V) and dI/dV(V), respectively, where details of the dI/dV(V) reveals a fine structure probably related to the Au surface states at c.a. -0.3 V [34,35]; a peak also appears at c.a. -0.6 V, which may be due to surface impurities. In general, the I(V) curves for bare Au and FTO surfaces over a range of -2 to 2 V display an ohmic behavior, with some fine details. In Fig. 8b, the I(V) spectrum presents the behavior of the jvdv1/Au-substrate. It is observed in Fig. 8b that the HOIS/Au sample does not exhibit significant ohmic behavior in the range of low voltages compared to the bare Au substrate, indicated by the flat region in (8b) defined by |V| < 0.5V, also observed before in similar experiments in ref. [36]. Finally, the apparent resistance, including that of the air gap, for the Au substrate is calculated to be about $2 \times 10^7$ Ω, while in (8b) with the deposited HOIS this apparent resistance increases tenfold due to the deposited jvdv1 film.

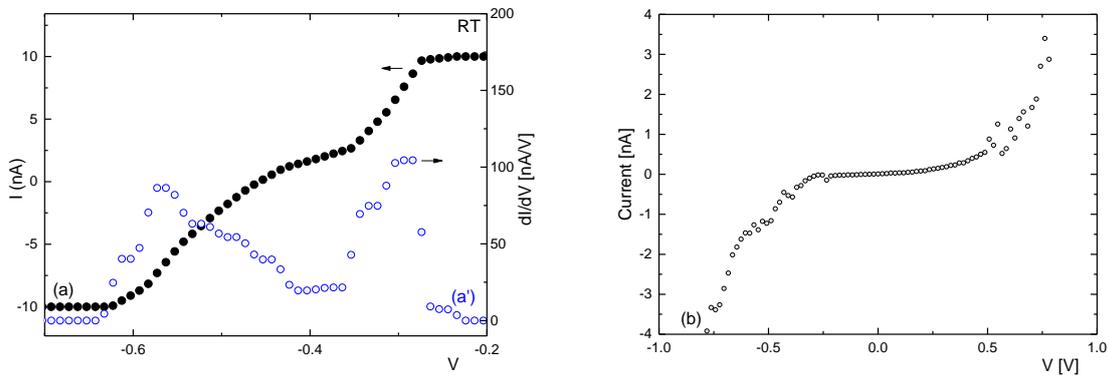

**Fig. 8.** STS spectroscopy results for the bare Au substrate (a,a') and the jvdv1 deposited on Au sample. Graphs are current (a,b) and dI/dV (a') as function of voltage. STM tip fixed with $V_{st}$=0.2 V; $I_{st}$=281 pA (left) and 103 pA (right).

On the contrary, when STS measurements were obtained for the jvdv1/FTO LED, an asymmetrical I(V) was observed, as shown in Fig. 9, where (a) and (b) curves refer to the forward and reverse voltage scans on the same point. The rectifying-like behavior was observed, when voltage was scanned from -5 up to 5V, as it was expected since the LEDs behave as diodes operating only at a specific polarity, which is when the tip is positive with respect to the sample and electrons are tunneling from the sample to the tip. This behavior was systematically observed for multiple spots on the sample. The data presented in Fig. 9 here, refer to the STM tip being initially fixed at a position defined with tunneling current set to 147 pA with the sample biased at -3.5V. The position of the tip at this high negative voltage should have been extended away from the sample, thus, it is natural to see that the measured conductance increases occurs at voltages lower than -3.5 V as observed at the left end of the graph in Fig. 9. However, when the voltage reaches 1.5 V, the forward current increases exponentially, which shows that even at large tip-sample distances the tunneling of the electrons toward the jvdv1/FTO is preferential. The reverse scan of the I-V, Fig. 9b, obtained when the voltage is swept in the opposite direction, yields a slightly shifted curve for positive voltages, but the shift depends on the position on the sample surface and may have to do with surface alteration or other polarization effects.

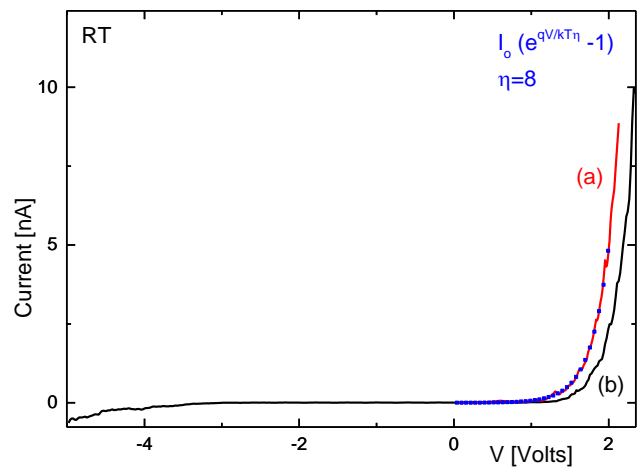

**Fig.9.** I-V characteristic of jvdv1 deposited on FTO as acquired with an STM. The red (a) and black (b) graphs represent the forward and reverse voltage scans. The dotted line represents a diode equation fitting for (a) with $\eta=8$.

The films used for LEDs or STS spectroscopy can be considered in general thick films in comparison to the thickness of the 2D inorganic layer as well as in comparison to the 2D platelets that appear to compose the films, as observed by SEM imagery. Therefore, the current conduction mechanism obviously is based on more effects other than a simple electron/hole drift throughout the volume of a single crystal, since the films presented here are polycrystalline. Moreover, the hole, electron and exciton diffusion lengths are important but these will be reported elsewhere. The physical explanation for the STS data can provide an insight on the electron transport mechanism. Therefore, it is important to discuss the absolute energy alignment of the conduction (CB) and valence (VB) bands of at least jvdv1, which is assumed to be close to that of the pristine 2D compounds.

In general, the underlying physics of the I-V graphs is quite complex as it involves the computation of pertinent data of the HOIS and the tip, such as work function, valence and conduction band edges, fermi level alignments and band bending at all interfaces. The absolute band edge positions have been computed here with the formalism as laid in ref. [37] for a similar 2D HOIS lead bromide based crystalline material. For comparison reasons and reference, this particular formalism provides for the well-known 3D $(CH_3NH_3)PbBr_3$ HOIS, CB and VB values of -3.11 eV and -5.42 eV, respectively, which are upshifted by 0.48 eV with the experimental values of the same 3D HOIS provided by ref. [38], thus, it is expected that the here computed 2D CB and VB band edge values should be downshifted by a similar value. CB and VB values for similar yet different 3D HOIS such as in ref. [39] were not considered although the latter are in qualitative agreement with those in [38]. Given the energy level alignment of the band edges of the tri-component device measured in the STM, Pt/Ir: $E_f$=-5.2 eV; ITO: -4.7 eV; HOIS: -3.86 eV CB and -7.62 VB,[38] it appears realistic to visualize electrons tunneling from the tip through various formed inter-platelet junctions to reach the FTO surface, where these recombine with tunneling holes into the jvdv1, assuming the HOIS is an isotropic material.

In this approach it is obvious that the Pt/Ir tip has an $E_f$ at c.a. -5.2 eV, while 2D $M_2PbBr_4$ compounds have a valence/conduction band edges at -7.62 eV and -3.86 eV, respectively. Therefore, it is natural to expect that a band bending voltage bias of 1.5 V will allow the tip electrons to drift/tunnel in the empty HOIS's conduction band. It is noted that in the approach here, a small negative bias has been used to obtain the STM images which reveal the occupied states of the surface [40]; for the STS measurements the tip had been retracted by large negative bias, therefore, the STS data yield lower current values. It is suggested, in summary, that the onset of the diodic behavior is in agreement with a single crystal approach towards the nature of the jvdv1 film, which means that the electrons/holes do travel from one nanoplatelet to a neighboring one. This hopping may not be the most efficient as the analysis of the relevant I(V) data of the first quadrant region of the graph in Fig. 9a, yields an ideality diode factor of 8. This was computed by fitting the simple diode equation for which no photovoltaic term was used since any stray light reaching the STM head was peaked at energies below the $E_g$ of the 2D HOIS.

This approach should also be realistic for the operation of the same device operated as LED with the exception of replacing the air-gapped Pt/Ir tip with Ga/In droplet electrode. However, the 2D HOIS or its DVs have a highly anisotropic 2D form, therefore, it is probable that the formation of defects as well as the close packing of nanoscopic 2D needles may allow, especially close to the FTO surface, electrons and holes to drift within the jvdv1 polycrystalline film by hopping from one 2D nanoplatelet to a neighboring one either via the intruding nanorods or by the sheer fact that the defects have forced the platelets to align randomly. Thus, such a random packing, which is partially due to the rather quick crystallization afforded by the heated substrate, may create routes for conduction that circumvent the non-conductive amine organic layers.

It is evident that the simple methodology in synthesizing HOIS and integrating them as LEDs presented in this work allows the fabrication of a blue excitonic LED operating at room temperature and it is expected that modification and advancements of this method will lead to even more surprising results and devices, but it is crucial that the amine used is of the fluorinated type as reported here.

As a final check on the mechanism that promotes EL in the defect only variants of jv221, while the latter does not show EL under any type of LED fabrication, we have also conducted time resolved PL measurements (TRPL) to elucidate the differences between all reported compounds. Other colleagues have found that the substrate temperature is crucial and is relevant to the crystallization of the 2D crystals as well as of their alignment to the substrate, while others have related the emitted EL (or loss of EL in the 2D HOIS) to the intricate relation of defects-carrier density-phonons and recombination time [41].

Here we support our suggestion, based on the TRPL data analysis, that the geometrical alignment of the 2D platelets is important for the single layer perovskite EL observation and that the excitonic recombination is not really assisted by the defects, but defects merely allow the 2D symmetry breaking, thus, allowing the drift of carriers [42]. The reason for that is that the PL decay spectra are mostly similar for the jv221 and its variants, while the decay times appear also to be quite similar. Thus, if the defects were forcing the recombination of the excitons that would have an impact on the decay lines with energies above, equal or below the 420 nm decay lines.

In Fig. 10, some of the TRPL spectra of the herein reported samples are presented, acquired at multiple PL lines, a.k.a 380, 400, 404, 410, 420, 450 and 500 nm. The relevant IRF spectra have not been depicted, for simplicity, yet they have been taken into consideration in the reconvolution algorithm analyzing the measured decay curves. In general, the reported decay curves show emission energy dependent decay curves, that cannot be uniquely modelled with a single or double exponential nor simple exponential stretched exponential factors[43] which would imply communication between localized states[44].

Fig. 10a presents a comparison of the jvdv1 TRPL spectra (0-50 ns) acquired at 400, 404, 410, 420 and 450 nm, whose PL peak is at 413-421 nm; thus, TRPL is been investigated at energies at the OA spectrum peak (410 nm), at slightly higher energies (400 and 404 nm), at the PL peak (420 nm) and at energies lower than that of the PL peak (450 nm). The emission energy dependence of the decay curves is evident in the log-log graphs. The jvdv1

TRPL decay spectra show a strong rapid intensity decrease up to 2 ns, with the fastest recombination attributed to the 420 nm line, followed by a second form of comparatively slower PL intensity decrease; this latter decrease is comparatively slower for the decay lines at 404 and 410 nm, implying a slow recombination path, also evidenced by the shoulder at the same curves at t=10 ns, probably due to some exciton localization mechanism. All jvdv1 observed emission lines exhibit a fast decay time of c.a. 200-400 ps while the prolonged delay appears to have a lifetime of c.a. 1-2 ns, in some cases fitted better with a stretched exponential of the form $e^{-(t/\tau)^\beta}$, with β = 0.48-0.66. However, it is obvious that the fastest recombination for jvdv1 is exhibited for the 420 nm line, which is also the only wavelength for which EL is being observed [45].

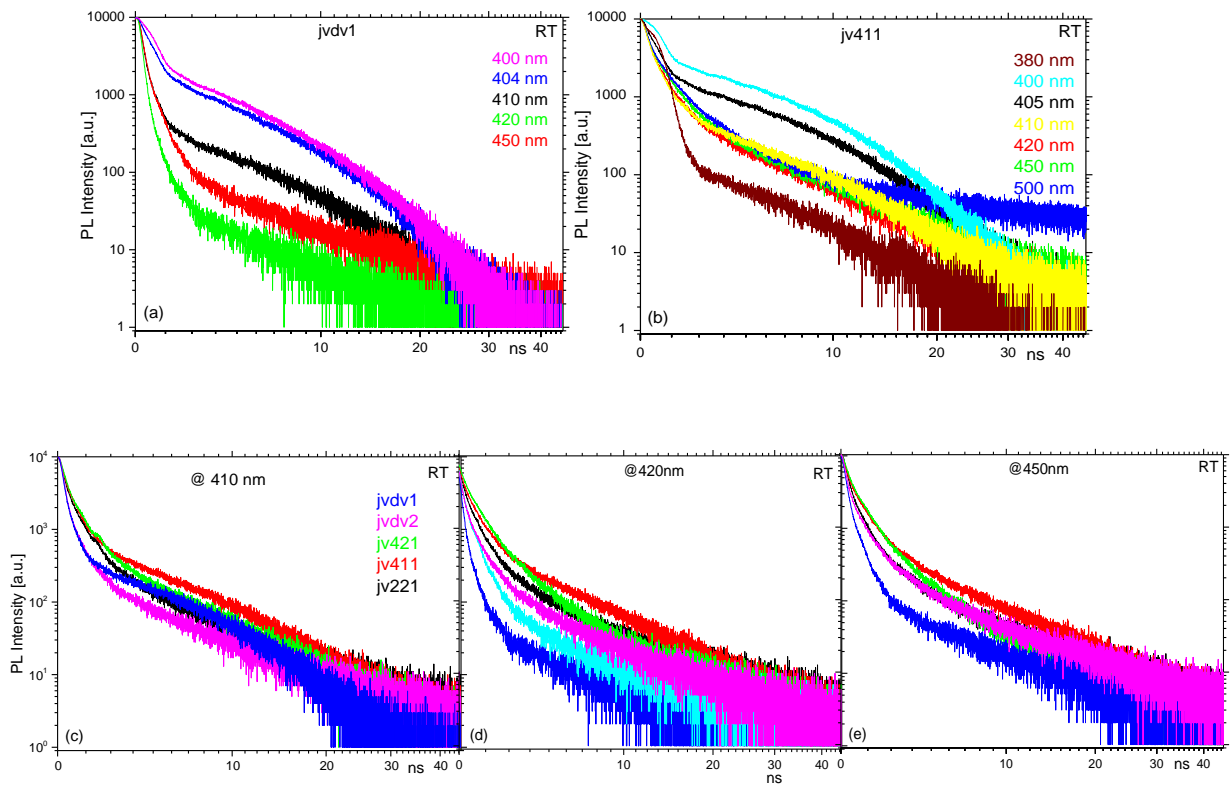

Fig. 10. RT TRPL spectra for (a) **jvdv1**, (b) jv411 for various emission lines; comparative TRPL spectra for all samples (c) at 410 nm, (d) at 420 nm and (e) at 450 nm. Colors applicable to {c,d,e}: **jvdv1** (blue), *jvdv2* (magenta), jv421 (green), jv411 (red), *jv221* (black).

Similar behavior is being observed for the jv411 sample (Fig.10b), where the high energy TRPL spectra (380, 400 and 405 nm) exhibit a delayed prolonged PL, as if there is an exciton localization, while the other emission lines appear to have a normal decay. All fitted parameters for the three important lines 410, 420 and 450 nm can be found in Table 2, where the model used was a double exponential out of which one was constrained to be normal while the other a stretched exponential. It is interesting that for jv411 the 420 nm emission line behaves as the 410 nm and 450 nm, thus, its PL and EL emission at 420 nm cannot be regarded as optimal in comparison to that of the 410 nm or 450 nm as far as the functionality of the defects, yet, jv411 also displays EL only at 420 nm. Sample jv411 has a decay line at 500 nm that shows PL at high times, thus, it can be inferred that there are a number of delayed recombination mechanisms that draw energy from the main emission at 420 nm.

Comparing the TRPL spectra among the different samples presented here, along with that of jv221 in Figs 10c,10d, 10e, for the 410, 420 and 450 nm lines, respectively, it can be immediately observed that jdv1 which does function as EL active material, shows recombination half times to be close to those of the jv221 for the all three tabulated emission line analyses. In fact, jvdv1 and the other variants, that function as EL materials, appear to have equivalent behavior as jv221 in terms of halftimes. Therefore, the effect that the defects exert on the electrically injected excitons' recombination time, is not correlated to the capability of have any of the reported semiconductors to behave as EL active materials[46]. It is, therefore, evident that the defects break the 2D platelet symmetry, which tend to align parallel to the surface, inducing common edge platelet points so that the electron hole pairs can diffuse as well as drift between the anode and the cathode.

**Table 2.** Parameters fitting the TRPL decay spectra: $I_0 + a_0 e^{-\left(\frac{t}{\tau_1}\right)^\alpha} + a_1 e^{-\left(\frac{t}{\tau_2}\right)^\beta}$ for the emission lines at 410, 420 and 450 nm, respectively, where $0 \leq I_o \leq 10$ and $I_{max}(t=0)=10^4$.

| Sample 410 \|\| 420 \|\| 450 nm | $\tau_1$ (ps) | $\tau_2$ (ns) | $\alpha$ | $\beta$ | $a_1/a_2$ |
|---|---|---|---|---|---|
| jvdv1 | 244 \|\| 314 \|\| 366 | 2.0 \|\| 1.0 \|\| 1.0 | 1 | 0.66 \|\| 0.57 \|\| 0.48 | 146 \|\| 119 \|\| 81 |
| jv411 | 431 \|\| 520 \|\| 637 | 1.8 \|\| 1.2 \|\| 1.1 | 1 | 1.80 \|\| 1.20 \|\| 1.10 | 20. \|\| 12. \|\| 7.0 |
| jv421 | 499 \|\| 556 \|\| 585 | 1.2 \|\| 1.3 \|\| 1.0 | 1 | 0.59 \|\| 0.74 \|\| 0.68 | 13 \|\| 10. \|\| 8. |
| jv221 | 446 \|\| 393 \|\| 375 | 1.2 \|\| 1.0 \|\| 1.0 | 1 | 0.62 \|\| 0.64 \|\| 0.65 | 19 \|\| 23. \|\| 20 |
| jvdv2 | 475 \|\| 501 \|\| 601 | 1.2 \|\| 1.0 \|\| 1.0 | 1 | 0.52 \|\| 0.57 \|\| 0.53 | 23 \|\| 18. \|\| 9. |

The PL quantum efficiency [47] has also been computed for the samples using a standard method, using solid quinine sulfate films and is being reported in Table 3. The computed values are in general higher than reported values for 3D materials [48], since the 2D excitonic confinement increases the oscillator strength, due to the fact that the spatial confinement of the electron and hole allow for effective recombination. It appears that the variant, jv421, which can function as EL material after careful deposition exhibits the highest EL efficiency probably due to the capping effect of the extra amine on the nanoparticles, by passivating any non-radiative mechanisms, while the pristine and the other defect variants have PLEQ that cannot be correlated to their capability to serve, in reported device fabrication scheme, as EL materials.

**Table 3.** PLQE values for the reported materials prepared on quartz substrates.

| Sample | PLQE (100%) |
|---|---|
| jvdv1 | 4.1 |
| jv411 | 5.8 |
| jv421 | 18.2 |
| jv221 | 10 |
| jvdv2 | 2.5 |

## 4. Conclusions

In conclusion, excitonic electroluminescence, observable by naked-eye at room temperature and at relatively low voltages, is observed for defect variants of a 2D hybrid organic-inorganic semiconductor, (4-fluorophenethylamine)$_2$PbBr$_4$, deposited as thin film between FTO anode and Ga/In cathode. The paradigm of the iodine pristine analogue that readily provided electroluminescent films, is now extended for the case of bromine, where the defects appear to be essential for the observation of EL. The pristine and its variants exhibit efficient excitonic optical absorption, photoluminescence and, cathodoluminescence, while color tuning can be achieved by incorporating different halide content. The defect's role is suggested to relate to the 2D platelet symmetry breaking, along the diffusion and drift of the carriers, although it also possible that the defects have an effect on the interfaces which govern the population and motion of electrons and holes. It is expected that this 4-fluorophenethylamine based class of 2D LD semiconductors will advance the modelling and application of 2D HOIS as it allows experimentation with simple, low cost devices.

**Conflicts of interest**

The authors declare that they have no conflict of interest.

**Acknowledgements**

This research did not receive any specific grant from funding agencies in the public, commercial or not-for-profit sectors.